%% file: author.tex
\documentclass[graybox]{svmult}

\usepackage{mathptmx}       % selects Times Roman as basic font
\usepackage{helvet}         % selects Helvetica as sans-serif font
\usepackage{courier}        % selects Courier as typewriter font
\usepackage{type1cm}        % activate if the above 3 fonts are
\usepackage{makeidx}         % allows index generation
\usepackage{graphicx}        % standard LaTeX graphics tool
\usepackage{multicol}        % used for the two-column index
\usepackage[bottom]{footmisc}% places footnotes at page bottom

\usepackage{algorithm}
\usepackage{algpseudocode}
\usepackage{amsmath}

\usepackage{enumerate}
\usepackage{marvosym}
\makeindex             % used for the subject index
\begin{document}

\title*{Learning automata based SVM for intrusion detection}
\author{Chong Di, Yu Su, Zhuoran Han, Shenghong Li}
\institute{Chong Di (\Letter) \at Shanghai Jiao Tong University, Shanghai, China, \email{dichong95@sjtu.edu.cn}
\and Yu Su \at Shanghai Jiao Tong University, Shanghai, China \email{suyu77@sjtu.edu.cn}
\and Zhuoran Han \at Shanghai Jiao Tong University, Shanghai, China \email{hzrtom@sjtu.edu.cn}
\and Shenghong Li \at Shanghai Jiao Tong University, Shanghai, China \email{shli@sjtu.edu.cn}}
\maketitle

%\abstract*{Each chapter should be preceded by an abstract (10--15 lines long) that summarizes the content. The abstract will appear \textit{online} at \url{www.SpringerLink.com} and be available with unrestricted access. This allows unregistered users to read the abstract as a teaser for the complete chapter. As a general rule the abstracts will not appear in the printed version of your book unless it is the style of your particular book or that of the series to which your book belongs.
%Please use the 'starred' version of the new Springer \texttt{abstract} command for typesetting the text of the online abstracts (cf. source file of this chapter template \texttt{abstract}) and include them with the source files of your manuscript. Use the plain \texttt{abstract} command if the abstract is also to appear in the printed version of the book.}

\abstract{As an indispensable defensive measure of network security, the intrusion detection is a process of monitoring the events occurring in a computer system or network and analyzing them for signs of possible incidents.
It is a classifier to judge the event is normal or malicious.
The information used for intrusion detection contains some redundant features which would increase the difficulty of training the classifier for intrusion detection and increase the time of making predictions.
To simplify the training process and improve the efficiency of the classifier, it is necessary to remove these dispensable features.
in this paper, we propose a novel LA-SVM scheme to automatically remove redundant features focusing on intrusion detection.
This is the first application of learning automata for solving dimension reduction problems.
The simulation results indicate that the LA-SVM scheme achieves a higher accuracy and is more efficient in making predictions compared with traditional SVM.
}
\section{Introduction}
\label{sec:1}
Recent years, with the development of internet and the rapid deployment of network applications, network security becomes an important research topic in the internet field.
As an indispensable defensive measure, the Intrusion Detection (ID) is a process of monitoring the events occurring in a computer system or network and analyzing them for signs of possible incidents \cite{guide}.
Specifically, the intrusion detection system is a classifier to judge the event is normal or malicious.
The computer system would record and store all the network logs whenever an event occurs.
Databases used for intrusion detection also include all of the information, such as DARPA Intrusion Detection Data Sets \footnote{http://www.ll.mit.edu/ideval/data/} and KDDCUP'99 \footnote{http://kdd.ics.uci.edu/databases/kddcup99/kddcup99.html}.
However, not all network logs and features could be used for intrusion detection.
Redundant and worthless information increases the difficulty when training the classifier for intrusion detection.
To simplify the training process and improve the efficiency of the classifier, it is necessary to remove redundant features from the training data.

As one of the most classic classification algorithms, Support Vector Machine (SVM) has been applied to predict attacks \cite{svm} and pretty superior results have been achieved.
Feature vectors containing too much redundant features make it knotty to solve the optimization problem in SVM and waste quiet a lot time to take the predict.
When using the SVM as the intrusion detection measure in network system, the efficiency of classifier is of great significance.
It is necessary to warn the manager as soon as possible whenever an intrusion occurs.

Learning automata is a reinforcement learning approach which chooses the optimal action from a set of actions through interacting with the random environment \cite{la}.
Taking the advantage of LA, in this paper, we propose a novel LA-SVM scheme to automatically remove redundant features focusing on intrusion detection problems.

The contributions of our work are summarized in the following:
\begin{itemize}
  \item 1. We present a learning automata based SVM scheme LA-SVM for intrusion detection.
  \item 2. This is the first application in dimension reduction \cite{dim-1} using learning automata. A novel automatic dimensionality reduction method is proposed, thus opening up a wide spectrum of research directions.
  \item 3. The simulation results indicate that the LA-SVM scheme can remove the redundant features successfully and select the most effective feature sets with even higher accuracy.
\end{itemize}

The rest of this paper is organized as follows.
In Section.\ref{sec:2}, we introduce the basic theoretical knowledge of SVM and LA which are the core compositions of our scheme.
We introduce the proposed LA-SVM scheme in Section.\ref{sec:3}.
The results of extensive simulations are presented in Section.\ref{sec:4}.
We conclude the paper in Section.\ref{sec:5}.

\section{Learning Automata}
\label{sec:2}
% Always give a unique label
% and use \ref{<label>} for cross-references
% and \cite{<label>} for bibliographic references
% use \sectionmark{}
% to alter or adjust the section heading in the running head
Learning automaton (LA) is a decision maker which can choose the optimal action and update its strategy through interacting with the random environment \cite{la}. As one of the most powerful tools in adaptive learning system, LA has a myriad of applications \cite{la-1}-\cite{la-3}.

\begin{figure}
  \centering\includegraphics[scale=.8]{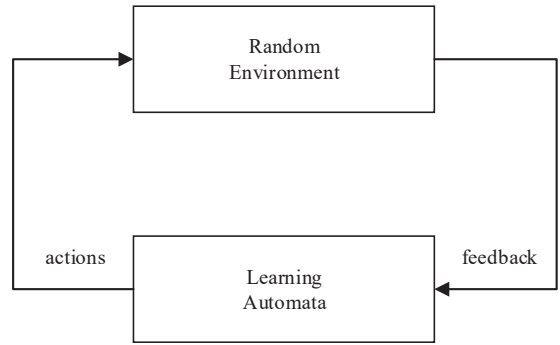}
\caption{Learning automata that interact with a random environment \cite{la}}
\label{fig:1}       % Give a unique label
\end{figure}

As illustrated in Fig.\ref{fig:1}, the process of learning is based on a learning loop involving two entities: the random environment and the LA. In this process, the LA continuously interacts with the random environment to get the feedback to its various actions. According to the responses to the various actions from the environment, LA will update the probability vector with a certain method. Finally, the LA attempts to learn the optimal action by interacting with the random environment through sufficient iterations.

A LA is defined by a quintuple $ < A,B,Q,F( \cdot , \cdot ),G( \cdot ) > $, where:

$\bullet$ $A = \{ {\alpha _1},{\alpha _2}, \cdots ,{\alpha _r}\} $ is the set of outputs or actions, and ${\alpha _t}$ is the action chosen by the automata at any time instant $t$.

$\bullet$ $B = \{ {\beta _1},{\beta _2}, \cdots ,{\beta _m}\} $ is the set of inputs to the automata, and ${\beta _t}$ is the input at any time instant $t$. The set $t$ could be finite or infinite. In this paper, we consider the case when $B = \{0,1\} $, where $\beta = 0$ represents the events that the LA has been penalized, and $\beta = 1$ represents the events that the LA has been rewarded.

$\bullet$ $Q = \{ {q_1},{q_2}, \cdots ,{q_s}\} $ is the set of finite states, and ${q_t}$ is the state of the automata at any time instant $t$.

$\bullet$ $F( \cdot , \cdot ):Q \times B \to Q$ is a mapping in terms of the state and input at any time instant $t$, such that, $q(t + 1) = F(q(t),\beta (t))$.

$\bullet$ $G( \cdot )$ is a mapping $G:Q \to A$, and is called the output function which determines the output of the automata depending on the state ${q_t}$, such that, $\alpha (t) = G(q(t))$.

The random environment interacted with LA is defined as $ < A,B,C > $, where $A$ and $B$ has been defined above. $C = \{ {c_1},{c_2}, \cdots ,{c_r}\} $ is the set of reward probability, and ${c_i}$ corresponds to an input action ${\alpha _t}$.

\section{Learning automata based SVM for intrusion detection}
\label{sec:3}
SVM is famous as a classic technique for solving a variety of classification and prediction problems.
Naturally, researchers have taken its advantages and apply the SVM directly to intrusion detection \cite{svm}.
Though the detection accuracy could meet the requirement to a certain degree.
Due to the negative impacts of redundant features, the algorithm is inefficient.

To overcome the drawbacks of traditional SVM using in intrusion prediction, in this paper, learning automata are exploited to remove unnecessary features automatically while we take SVM as a basic classifier.
Firstly, the problem of intrusion detection is introduced. Then we describe the learning automata problem mapping. At last, the procedure of proposed LA-SVM scheme is presented.

\subsection{Problem formulation and Solution constrction}
Given a processed network log $F = \{f_{1}, f_{2}, ..., f_{N}\}$ which contains a series of features, where $N$ is the number of features.
The processed network logs correspond to events one by one.
The purpose of intrusion detection is to judge the event is normal or malicious by analyzing the corresponding features.
And if it is malicious, distinguish the type of attacks and warn the managers as soon as possible.

The learning automata mapping includes two main entities: a learning automaton and a random environment.

\emph{1) Learning automata perspective:} In the LA-SVM scheme, the entirety of features is modeled as a self-update learning automaton and each feature $f_{i}$ in feature set $F$ is considered as an action of the learning automaton.
The structure of the learning automaton in LA-SVM scheme could be described by $\{\alpha, \beta, P\}$, where
\begin{itemize}
  \item $\alpha = \{\alpha_{1}, \alpha_{2}, ..., \alpha_{N}\}$ is the action set, which corresponds to the set of features $F$. Each action is mapped to a feature and $N$ is the number of actions.
  \item $\beta = \{0, 1\}$ is the feedback from the random environment, where $0$ corresponds to reward and $1$ corresponds to penalty.
  \item $P = \{p_1, p_2, ..., p_N\}$ is the action probability vector. At each time instant $t$ during the learning process of LA-SVM, we choose an action randmly according to the probability distribution $P$ which is initialized to uniform distribution. Thus, at $t = 0$, $P(t) = \{1/N,...,1/N\}$.
\end{itemize}

\emph{2) Environment perspective:} The random environment decides the value of the feedback $\beta_t$ at every time instant $t$. We will discuss the standard of reward and penalty in next subsection.

\subsection{The LA-SVM algorithm}

Before presenting the LA-SVM scheme, we would explain the symbols used in our algorithm first.
The action set $A$ consists of all the features, and the result of the scheme is to select the indispensable features and remove all of the redundant features.
The training data $TRAIN\_DATA$ is divided into training part and validation part.
Note the training subset as $Tr$ and the validation subset as $Val$.
The testing data $TEST\_DATA$ is used to examine final effect of the algorithm.

$T_1$ and $T_2$ are two thresholds. $T_1$ is the lower limit of accuracy and $T_2$ is the lower limit of choice probability.
$\Delta$ is the smallest step size. $R$ is the removed-feature set which is initialized to an empty set.

And now, we will present the procedure of LA-SVM scheme.

\begin{algorithm}[htb]
\caption{ Learning automata based SVM for intrusion detection }
\label{alg:LA-SVM}
\begin{algorithmic}[1]
  \State Initialize $T_1$, $T_2$, $\Delta$;
  \State \textbf{Repeat}:
  \State At time instant $t$, select an action $\alpha(t) = \alpha_{i}$, according to the probability distribution $P(t)$;
  \State Choose a training subset $Tr(t)$ and a validation subset $Val(t)$ randly from the subsets and ensure that $Tr(t) \not= Val(t)$;
  \State Remove the $i-th$ feature from training subset $Tr(t)$ and a validation subset $Val(t)$ temporarily;
  \State Train the SVM using the training subset $Tr(t)$ and then get the trained model $model(t)$;
  \State Use the validation subset $Val(t)$ to test the $model(t)$, and get the accuracy $accuracy(t)$;
  \State if $accuracy(t) >= T_1$, the random environment feed back a reward, which means $\beta(t) = 0$:
  \State \indent Update the probability vector $P(t)$ according to the following equations:
  \State \indent \indent $p_{j}(t) = max\{p_{j}(t)-\Delta, 0\},\forall j \not= i $;
  \State \indent \indent $p_{i}(t) = min\{1-\sum_{j\not=i}{p_{j}(t)\}, 1}$;
  \State if $max(P(t) >= T_2$:
  \State \indent $[m, p_m] = max(P(t)$, where $m$ corresponds to the action with highest probability;
  \State \indent Remove the $m-th$ feature from all of the subsets permanently and add this feature to the removed-feature set $R$;
  \State \indent Reinitialize the action set $A$ and probability vector $P$;
  \State \textbf{Until} can not meet the convergence conditions;
  \State Remove all of the feature in the removed-feature set $R$ from training data $TRAIN\_DATA$ and testing data $TEST\_DATA$;
  \State Train the SVM using the processed training set $TRAIN\_DATA$ and then get the final model;
  \State Use the final model to make the predict the label of processed testing data $TEST\_DATA$.
\end{algorithmic}
\end{algorithm}

The core process of LA-SVM could be summarized as removing the redundant features one by one.
Before the iteration process begins, we will randomly select the training subset $Tr$ and the validation subset $Val$ for $r$ times.
Afterwards, we would train SVM and test the trained models using these subsets for $r$ times.
Then, the threshold $T_1$ is initialized to the average accuracy, where $T_1=\sum\limits_{i = 1}^r {accuracy_{i}}$.
At each time instant $t$, we select an action $\alpha(t) = \alpha_{i}$, according to the probability distribution $P(t)$.
Then we temporarily remove the feature corresponding to the action $\alpha_{i}$ from the randomly selected training subset and validation subset.
The processed training and validation subset are used to train a SVM model and evaluate performance of trained model according to the classification accuracy.
If the accuracy is higher than the initialized threshold $T_1$, it indicates that the removed feature may be redundant in intrusion detection.
In this context, the random environment would feed back a reward to learning automaton.
Whenever the learning automaton gets a feedback reward, we will update the action probability vector $P$ using the following formula which means that the probability $p_i$ corresponding to the action $\alpha_i$ will be increased and others are increased.
\begin{equation}
p_{j}(t) = max\{p_{j}(t)-\Delta, 0\},\forall j \not= i
\end{equation}
\begin{equation}
p_{i}(t) = min\{1-\sum_{j\not=i}{p_{j}(t), 1}\}
\end{equation}

Through this way, the probabilities of redundant features will be higer than the probabilities of necessary features.
Thus, the redundant features are more likely to be selected and get more chance to be evaluate by the random environment in the next iteration process.
Whenever the probability of an action $\alpha_m$ is higher than the threshold $T_2$, the corresponding feature is considered to be redundant enough.
After removing the corresponding feature from training data and reinitialize the related vectors, the redundant features would be found out one by one until there is no feature could be removed.

\section{Simulation}
\label{sec:4}
In this section, the performance of the proposed LA-SVM scheme is evaluated from two aspects by comparing to the SVM scheme.
The first evaluation standard is the classification accuracy.
The most important precondition of dimensionality reduction is that it does not reduce the accuracy.
The second standard is the time that the trained model used to make predictions.

\subsection{Data specification and data preprocessing}
The $KDD Cup 1999$ dataset contains a standard set of data to be audited, which includes a wide variety of intrusions simulated in a military network environment.
The dataset consists of 41 features and a label.

The features could be classified into four different categories, basic features, contents features time based traffic features and host based traffic features.
The data type of most features is continuous.
The values of some features like $land$, $logged_in$ can be 0 or 1.
We treat them in the same way as continuous features.
The $protocol_type$ feature has three different value corresponding to number 1, 2, and 3.
The $flag$ feature has 11 different values and the $service$ feature has 66 distinct values.
We use clustering algorithm proposed by $Hernndez-Pereira$ \cite{cluster} to reduce the dimensionality before transforming these different values into numbers.
At last, we perform necessary scaling to normalize the data.

The label specifies the event is normal or malicious.
There are different types of attacks and we classify them into 4 categories, including denial of service(DoS), User to root(U2R), Remote to local(R2L), Probing(PROBE).
Thus, in our experiments, the label has five different values $Label = \{0,1,2,3,4,5\}$, where 0 corresponds to normal and others correspond to different attacks.

We use $kddcup.data\_10\_percent$ dataset as the trarning data, which contains $204743$ normal events, $283993$ DoS attacks, $52$ U2R attacks, $1126$ R2L attacks and $4107$ PROBE attacks.
To improve the typicalness of subsets, at each time instant $t$, we randomly select $5000$ normal events and $5000$ DoS attacks combined with all of the other attacks to form a training subset or a validation subset.
Dataset $kddcup.corrected\_labels$ is used as a whole to evaluate the performance of scheme.

\subsection{Evaluation results}
Before presenting the simulation results, we will show the values of related parameters.
We set the threshold $T_2 = 0.8$.
The smallest step size $\Delta = 1/N/10 = 0.00244$.
As mentioned above, We calculate the threshold $T_1$, where $\sum\limits_{i = 1}^r {accuracy_{i}} = 74.2827\%$.
We set $T_1 = 74.2827\%$ in experiment LA-SVM-1 to LA-SVM-2 and set $T_1 = 74.2827\%+20\% = 94.2827\%$ in experiment LA-SVM-3 to LA-SVM-5.

The final features selected by LA-SVM scheme, the accuracy and the time used to making predictions are presented in Table.1.
%%table 1
\begin{table}[!hpb]
  \centering
  \caption{ Performance comparison of SVM and LA-SVM }
  \label{table:1}
  \begin{tabular}{l|l|l|l|l|l|l}
  \hline
  & SVM &  LA-SVM-1 &  LA-SVM-2 & LA-SVM-3 &  LA-SVM-4 &LA-SVM-5\\ \hline
number of features & 41 & 5 & 3 & 5 & 4 & 4\\ \hline
features & 1-41 & 4,5,12,24,26 & 2,5,24 & 4,5,6,24,25 & 2,5,6,24 & 3,5,6,24\\ \hline
accuracy &  86.1383\% & 85.96\% & 85.83\%	& 96.70\% & 96.00\%	& 96.13\% \\ \hline
test\_time(s) & $0.228174$ & $0.060402$ & $0.055509$	& $0.069832$ & $0.057892$	& $0.057191$ \\ \hline
  \end{tabular}

  $^a$ the feature names corresponding to the features:
   $2-protocol\_type,3-service,4-flag,5-src\_src\_bytes,24-srv\_count,25-serror\_rate,26-src\_error\_rate$
\end{table}

The simulation results indicate that the proposed LA-SVM scheme could achieve the equivalent accuracy as the SVM scheme when we set the parameter $T_1 = \sum\limits_{i = 1}^r {accuracy_{i}}$.
And when we further optimize the parameters, set $T_1 = \sum\limits_{i = 1}^r {accuracy_{i}}+20\%$, the accuracy would even be higher.
It reduces the feature dimension from 41 to less than 5 and improves the accuracy from 86\% to more than 96\%.
Thus, the LA-SVM is a successful dimension reduction method for intrusion detection.
Furthermore, the time used to predict the event is also greatly reduced, which means the scheme would be more efficient in practical application.

\section{Conclusion}
\label{sec:5}
In this paper, we propose a novel dimension reduction method LA-SVM for intrusion detection.
Compared to traditional SVM, it simplifies the complexity of optimization problem when training the classification model.
The simulation results indicate that the LA-SVM scheme achieves a higher accuracy and is more efficient in making predictions.
In addition, this is the first application of learning automata for solving dimension reduction problems.
Our further works aim to establish a theoretical model of learning automata based dimension reduction method and exploit it to solve other problems.

\begin{acknowledgement}
  This research work is funded by the State Grid Corporation of China (SGCC) Science and Technology Project (SGRIXTKJ［2017］133), the National Key Research and Development Project of China (2016YFB0801003), and the Key Laboratory for Shanghai Integrated Information Security Management Technology Research.
\end{acknowledgement}

\input{referenc}

\end{document}

%% file: referenc.tex
%%%%%%%%%%%%%%%%%%%%%%%% referenc.tex %%%%%%%%%%%%%%%%%%%%%%%%%%%%%%
% sample references
% %
% Use this file as a template for your own input.
%
%%%%%%%%%%%%%%%%%%%%%%%% Springer-Verlag %%%%%%%%%%%%%%%%%%%%%%%%%%
%
% BibTeX users please use
% \bibliographystyle{}
% \bibliography{}
%